\documentclass[utf8]{FrontiersinHarvard} 

\usepackage{url,hyperref,lineno,microtype,subcaption}
\usepackage[onehalfspacing]{setspace}

\linenumbers

\def\keyFont{\fontsize{8}{11}\helveticabold }
\def\firstAuthorLast{Yigit {et~al.}} 
\def\Authors{Zehra Yigit\,$^{1,*}$, Ertugrul Basar\,$^{2}$ and Ibrahim Altunbas\,$^{1}$}
\def\Address{$^{1}$\mbox{Department of Electronics and Communication Engineering, Istanbul Technical University},  Maslak 34469, Istanbul, Turkey. \\
	$^{2}$\mbox{Communications Research and Innovation Laboratory (\text{CoreLab}),  Department of  Electrical and} Electronics Engineering, Ko\c{c} University, \mbox{Sariyer 34450, Istanbul, Turkey.}  }

\def\corrAuthor{Zehra Yigit}

\def\corrEmail{yigitz@itu.edu.tr}

\begin{document}
	\nolinenumbers
	\onecolumn
	\firstpage{1}
	
	\title[Over-the-Air Beamforming with RIS]{Over-the-Air Beamforming with Reconfigurable Intelligent Surfaces} 
	
	\author[\firstAuthorLast ]{\Authors} 
	\address{} 
	\correspondance{} 
	
	\extraAuth{}

	\maketitle

	\begin{abstract}
		
		{	Reconfigurable intelligent surface (RIS)-empowered communication} is a  revolutionary technology that enables to manipulate wireless propagation environment via smartly controllable  low-cost reflecting surfaces. However, in order to outperform conventional communication systems,  an RIS-aided system  with   solely passive reflection requires an extremely large surface. To meet this challenge, the concept of active RIS, which performs simultaneous amplification and reflection  on the incident signal at the expense of additional power consumption,  has been recently introduced. In this paper, deploying an active RIS, we propose a novel beamforming concept,  \textit{over-the-air beamforming}, for RIS-aided {multi-user multiple-input single-output (MISO)} transmission schemes without requiring any pre/post signal processing hardware designs at the transmitter and  receiver sides. In the proposed {over-the-air beamforming-based} transmission scheme, the  reflection coefficients of the active RIS elements are customized to  maximize the sum-rate gain. To tackle this issue, first, a non-convex quadratically constrained quadratic programming (QCQP) problem is formulated. Then, using semidefinite relaxation (SDR) approach, this optimization problem is converted to  a convex feasibility problem, which is efficiently solved using the CVX  optimization toolbox. Moreover, taking inspiration from  this  beamforming technique,  a novel high-rate receive index modulation (IM) scheme with a low-complexity sub-optimal detector is developed. Through comprehensive simulation results, the sum-rate and bit error rate (BER) performance of the proposed designs are investigated. 

		\tiny
		\keyFont{ \section{Keywords:} 	Reconfigurable intelligent surface (RIS),  active RIS, over-the-air beamforming, multi-user (MU) transmission,  index modulation (IM).} 
	\end{abstract}
	
	\section{Introduction}
	
	Customizing propagation environment via reconfigurable intelligent surfaces (RISs) has been  an appealing field for  wireless communication and provides novel insights about future generation networks.  These light-weight and  cost-effective electronic elements have been  regarded as a game changer technology for  conventional communication systems with power-hungry and complex hardware designs \citep{basar2019wireless}. Particularly,  RISs are programmable metasurfaces that  are capable of configuring the { propagation environment} in a desired manner via performing reflection, amplification, absorption, refraction, etc. \citep{di2020smart}.  However, the most of the extant literature particularly focuses on the  application of the RIS with passive reflection in various emerging systems \citep{basar2019wireless,di2020smart,gong2020toward}.

	In the early studies,  a passive RIS is deployed  for enhancing  transmit  signal quality of   single-antenna \citep{basar2019transmission} and  multiple-antenna systems \citep{yu2019miso,zhang2020robust,yigit2020low}. In  subsequent studies,
	an   RIS is facilitated for numerous objectives of single-user and  multi-user  systems, such as  promoting energy-efficiency  \citep{huang2019reconfigurable,bjornson2019intelligent}, enhancing error performance \citep{ye2020joint,ferreira2020bit} and improving achievable rate  \citep{zhang2020capacity, perovic2021achievable,di2020hybrid}. Further, novel deep learning-based solutions for  passive RIS designs \citep{taha2021enabling,kundu2021channel} and security enhanced RIS-aided communication systems {\citep{shen2019secrecy}}, \citep{ almohamad2020smart,dong2020secure}   are proposed. On the other hand,  index modulation (IM)  principle, which is  emerged as a promising energy-efficient solution to meet high data-rate demand of future wireless networks  \citep{basar2017index}, is  beneficially amalgamated into  the RIS-empowered communication \citep{basar2020reconfigurable,li2021single}. {Considering   more conventional IM designs,  \citep{li2021single, basar2020reconfigurable} put forward   RIS-aided receive IM schemes, which maximize the signal powers of the target receive antennas}.  However, in \citep{guo2020reflecting,lin2020reconfigurable},  novel reflection modulation (RM) concepts, which innovatively utilize the RISs for delivering additional information, are proposed. Above all, main limitation of the aforementioned studies is the lack of comprehensive practical  insights on considered system configurations.  Towards this aim, { a low-complexity joint beamforming optimization that considers  the effect of hardware impairments on the  performance of RIS-aided multi-antenna systems are investigated in \citep{shen2020beamforming}},    different RIS prototypes are introduced  for real-time implementations in  \citep{dai2020reconfigurable,tang2020mimo}, and  realistic physical channel models for millimeter-wave (mmWave) \citep{basar2021indoor} and sub-$6$ GHz bands \citep{kilinc2021physical,yigit2021simmbm} are presented. Nevertheless,  the abovementioned  system designs  suffer from the multiplicative  path attenuation due to the inherent drawback of the RIS-aided designs, and  achieve negligible performance gains over the conventional communication systems.

	Recently, to tackle above challenges, the concept of active RIS, which performs simultaneous amplification and reflection  on the incident wave, is introduced in \citep{zhang2021active, long2021active}. Accordingly,  the magnitudes and the phases of the reflecting elements of the active RIS, which are equipped with   additional power amplifiers, are properly tuned  in a customized way \citep{basar2021present}. Therefore, at the cost of additional power consumption, active RIS-aided systems are capable of  achieving enhanced capacity gains \citep{long2021active}.  In a recent  study on designing active RISs, via leveraging  power amplifiers and radio frequency (RF) chains  \citep{nguyen2022hybrid},  dynamic and fixed hybrid  RIS architectures are constructed. Further, for improving the data rate, a new RM design,
	which employs  the sub-groups of a hybrid RIS as information transfer units, is presented in \citep{yigit2021hybrid}. In follow-up studies, the concept of the active RIS is deployed for  beamforming optimization of the RIS-aided multi-user systems \citep{gao2022beamforming,thanh2022hybrid}. Above all, the potential of the active RIS-aided systems for achieving  enormous performance gains will enable   to  develop promising solutions for future research.


	In this study, unlike the conventional precoding techniques that employ power-hungry and hardware-complex devices \citep{sohrabi2016hybrid},  for  RIS-aided  multi-user downlink transmission systems, we propose  a novel  \textit{over-the-air beamforming}  technique with the aid of an active RIS  to exploit its capability of manipulating the magnitude of the incident wave. In other words, the main motivation of the over-the-air beamforming scheme is to simplify the transmitter and receiver ends of the overall network while transferring inter-user interference elimination tasks completely to an active RIS. {Therefore, this paper proposes two novel over-the-air beamforming schemes that mitigate the burden of signal processing on the transmitter and receiver sides. }  In the proposed over-the-air beamforming-based  transmission scheme, it is assumed that a multi-antenna transmitter serves $K$  single-antenna users through an active RIS without utilizing any other signal processing tasks at the transmitter and the receiver sides.  Then, the reflection coefficients of the  active RIS is properly adjusted to maximize the sum-rate of the overall system. 
	 {Moreover, taking inspiration from this over-the-air beamforming concept, a new
	receive IM scheme that transmits additional information bits to specify the index of
	the effective received antenna is also proposed.} 
	  Contrary to the traditional receive IM systems \citep{luo2021spatial, zhang2013generalised},   in the proposed  system, since no precoding is applied at the transmitter, the reflection coefficients of the active RIS are rectified to steer the incident signal into the intended receive antenna.  On the other  hand, since the receive IM scheme benefits from the multi-antenna transmission at the user side and IM system design at the receiver side,  it shows the favourable features of both, such as high  spectral efficiency and improved performance. {In these proposed over-the-air downlink beamforming  and over-the-air  uplink receive IM schemes, to optimize
	  	 reflection coefficients of the active RISs,  two distinct semidefinite relaxation
	  	(SDR)-based optimization problems are formulated, which can be effectively solved
	  	through the CVX convex optimization toolbox \citep{gb08}.  } Furthermore,   the achievable rate and bit error rate (BER) performance of the proposed over-the-air beamforming-based transmission schemes are investigated through extensive computer simulations. 
	
	The rest of the paper is organized as follows. In Section 2, after giving a short review of  the conventional zero-forcing (ZF) precoding, we introduce the system model of the proposed over-the-air beamforming-based multi-user multi-antenna transmission scheme.   In Section 3,  the  over-the-air beamforming-based receive IM scheme and its low-complexity receiver detection  are introduced. Section 4  provides the achievable rate and BER results of the proposed over-the-air beamforming based transmission systems, and the conclusions are drawn in Section 5.
	
	\textit{Notations:}  Throughout this paper,  matrices  and vectors are  denoted by  boldface upper-case and boldface lower-case letters, respectively. $(\cdot)^{\mathrm{T}}$ represents   transpose and   $(\cdot)^{\mathrm{H}}$ denotes the Hermitian transpose operation.    $\left\|\cdot \right\| $, $\mathrm{rank}(\cdot)$, $\mathrm{Tr}(\cdot)$  and $\mathrm{diag}(\cdot)$ are stand for rank, trace and diagonalization of a matrix, respectively.  Absolute value of a scalar is denoted by  $|\cdot|$, while $\circ$ represents the Hadamard product.  $\mathbb{E}\left\lbrace \cdot\right\rbrace $ is used for expectation and $\mathcal{CN}(\mu,\sigma^2)$ represents a complex Gaussian random variable with $\mu$ mean and $\sigma^2$ variance.  $\mathbf{I}$ stands for the identity matrix, while $\mathcal{O}(\cdot) $ denotes big $\mathcal{O}$ notation. 

	

	\section{Over-the-Air Beamforming with  RIS }  
	
	In this section, after a review of conventional transmit precoding,  the  over-the-air  beamforming concept is introduced for {multi-user multiple-input single-output (MISO)}  downlink transmission systems.
	
	\subsection{ Conventional Transmit  Precoding}
	
	Considering a typical multi-user downlink transmission system without an RIS,  {a base station (BS)} transmitter (T) with $T_x$ antennas  is assumed to perform  ZF precoding to alleviate  interference between  $K$ single-antenna users \citep{spencer2004zero}. Let $\mathbf{F}\in\mathbb{C}^{K\times T_x}=\sqrt{L_{D}}\bar{\mathbf{F}}$ represents the channel matrix of the direct links between the T and the users,   where  
	$\bar{\mathbf{F}}\in\mathbb{C}^{K\times T_x}$ is modeled as independent Rayleigh fading channel matrix with $\sim\mathcal{CN}(\mathbf{0},\mathbf{I})$ and $L_{{D}}$ is the corresponding  path attenuation, which is calculated as $L_{D}=C_0d_{D}^{-\beta_{D}}$, where $C_0$ is the reference path attenuation at a distance of $1$ meter (m) and  $d_{D}$ is the distance between  T and the users. Then, 
	 the received signal of the  $k$-th user ($\text{U}_k$), for  $k\in\left\lbrace 1,2,\cdots,K\right\rbrace$, becomes
	\begin{equation}
		y_k=\mathbf{f}_k\mathbf{w}_k^{\mathrm{H}}x_k+\sum_{i\neq k}^{K}\mathbf{f}_k\mathbf{w}_i^{\mathrm{H}}x_i+{c}_k
	\end{equation}
	where  $x_k$ being an $M$-ary phase shift keying  (PSK) signal to be transmitted over the $k$-th transmit antenna. Here,   $\mathbf{f}_k\in\mathbb{C}^{1\times T_x}$ is the $k$-th row of the channel matrix $\mathbf{F}$ corresponding to the channel vector between T-$\text{U}_k$,  $\mathbf{w}_k\in\mathbb{C}^{1\times T_x}$ is the precoding vector for $\text{U}_k$ and $c_k\sim\mathcal{CN}(0,\sigma_s^2)$ is the static noise at $\text{U}_k$.  
	Therefore, the signal-to-interference-plus-noise-ratio (SINR) at $\text{U}_k$ can be calculated as
	\begin{equation}
		\gamma_k=\frac{\left\| \mathbf{f}_k\mathbf{w}_k^{\mathrm{H}}\right\|^2}{\sum_{i\neq k}^{K}\left\| \mathbf{f}_k\mathbf{w}_i^{\mathrm{H}}\right\|^2+\sigma_s^2 }.
	\end{equation}
	Moreover, the overall transmit ZF  precoding matrix, exploiting the perfect channel state information (CSI), can be obtained  as \citep{spencer2004zero}
	\begin{equation}
		{\mathbf{W}}=\sqrt{\zeta}(\mathbf{F}^{\mathrm{H}}\mathbf{F})^{-1}\mathbf{F}^{\mathrm{H}}.
	\end{equation}
	where $\mathbf{W}\in\mathbb{C}^{T_x\times K}=[\mathbf{w}_1^{\mathrm{H}},\cdots, \mathbf{w}_K^{\mathrm{H}}]$ and $\zeta$ is a scaling  constant  to meet the total power constraint $P_{T}$, such that   $\mathbb{E}\left\lbrace \mathbf{W}\mathbf{W}^{\mathrm{H}}\right\rbrace=P_{T} $.

	\subsection{System Model of Over-the-Air Beamforming with  RIS }
	In this subsection, {after a brief introduction of the active RIS concept}, the system model of the over-the-air beamforming-based multi-user transmission system is introduced.

\subsubsection{Active RIS}
{The principal drawback of   RIS-aided communication systems is the  inherent multiplicative  path attenuation along the RIS-aided indirect link, which is hardly compensated by the RIS with passive reflecting elements \citep{zhang2021active,basar2021present}. Therefore, to overcome this challenge, an RIS architecture with active reflecting elements that enable to configure both the magnitude and phase of the incident wave at the expense of an additional power consumption, is  recently proposed \citep{zhang2021active,khoshafa2021active}. Therefore, unlike the passive RISs,  the  active RISs reflect incident signal with amplification via employing additional  power  circuitry.    Although the active reflecting elements have a similar capability of amplifying the incident signal as in the {full-duplex} amplify-and-forward (AF) relays, their hardware constructions are completely different from each other. While the AF relays embody a circuitry for amplification in their hardware constructions and  they  are also  externally equipped with high power-consuming RF chains to transmit and receive signals \citep{wu2019intelligent},  the active reflecting elements employ reflective-type power amplifiers to simultaneously  rectify the magnitude and phase of the incident wave \citep{zhang2021active}.    }
\begin{figure}[htp!]
	\begin{center}
		\includegraphics[width=0.8\linewidth]{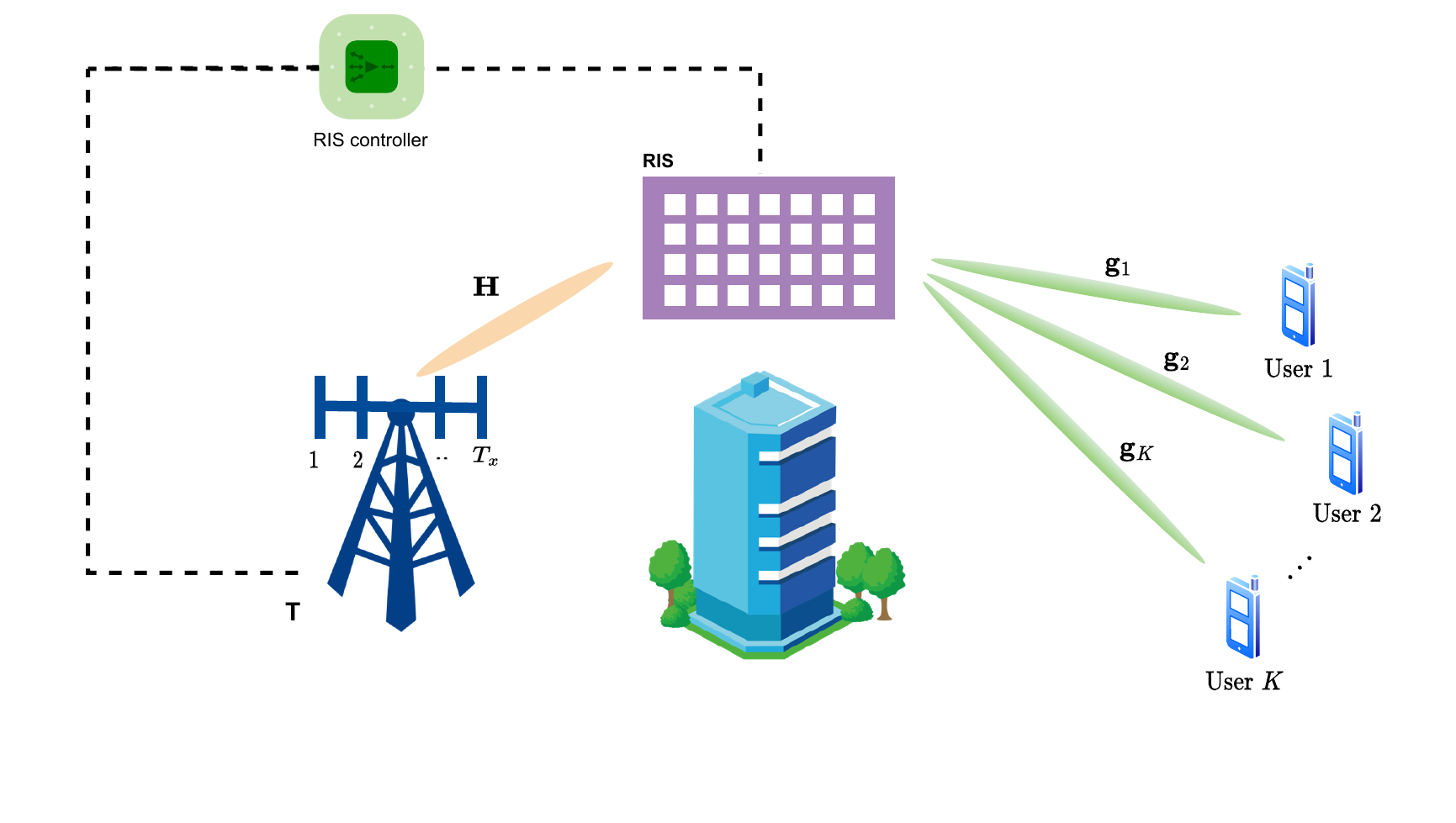}
	\end{center}
	\caption{ Over-the-air beamforming-based multi-user downlink transmission system.}\label{sys1}
\end{figure}
\subsubsection{System Model}
	An overwhelming literature on passive RIS-aided multi-user transmission  deploys the RIS as a passive beamformer after a preprocessing is conducted at the transmitter \citep{wu2019intelligent,yan2020passive}.  { However, in the proposed over-the-air beamforming concept,   to avoid power-hungry hardware constructions at the transmitter and the users,    exploiting simultaneous amplification and reflection capabilities of the active reflecting elements,  both  active and  passive beamforming are carried out at an active RIS.  Accordingly, the reflection coefficients of the active RIS are optimized to maximize the achievable rate of the overall system.}
	
	As given in Figure \ref{sys1}, in the proposed scheme,   the direct  transmission links between  T with $T_x$  antennas  and $K$ single-antenna users  are   neglected due to obstacles, thus,    the communication is established through an active RIS with $N$ reflecting elements.  In the proposed over-the-air  beamforming-based multi-user transmission, it is assumed that T and the users have the perfect CSI about T-RIS and RIS-users channels, which is conveyed to a  smart RIS controller via a feedback control link \citep{wu2019intelligent}.    Moreover, at the transmitter side, without {requiring} any additional  signal processing approaches for interference mitigation,   the overall  signal is conveyed to the users through the RIS. 
	{Hence, }    unlike the traditional beamforming techniques that employ complex and  power-hungry  signal processing hardware \citep{sohrabi2016hybrid,el2014spatially}, the RIS  is designed as a  beamformer to alleviate multi-user interference by adjusting the amplitude and phase of each reflecting element. Towards this aim, the  RIS elements are assumed to be equipped with additional power circuitry to  modify both the magnitude and the phase of the incident signal \citep{zhang2021active,nguyen2022hybrid,yigit2021hybrid}.   {Furthermore, in the proposed system, since all transmit antennas  simultaneously convey their own  $M$-PSK modulated signals, a spectral efficiency of $\eta_{\text{MU}}=T_x\log_2(M)$ [bits/s/Hz] is achieved}.
	
	Let us assume that the channels between T-RIS are presented by the matrix  $\mathbf{H}\in\mathbb{C}^{N\times T_x}=\sqrt{L_{{T}}}\bar{\mathbf{H}}$ and $\mathbf{g}_k\in\mathbf{C}^{1\times N}=\sqrt{L_k}\bar{\mathbf{g}}_k$ represents the  vector of channel coefficients between the RIS and  $\text{U}_k$, where $L_{T}$ and $L_k$ correspond to path attenuation between T-RIS and RIS-$\text{U}_k$ links  for $k\in\left\lbrace 1,2,\cdots,K \right\rbrace $, respectively.  Here, for $d_T$ and $d_k$ being the corresponding distances,  using a well-known distance-dependent model, the path attenuations are obtained as $L_T=C_0d_T^{-\beta_T}$ and  $L_k=C_0d_k^{-\beta_k}$, where  $\beta_T$ and $\beta_k$ are the path loss exponents at T-RIS and RIS-$\text{U}_k$, respectively. In the proposed system, the matrix $\bar{\mathbf{H}}\in\mathbb{C}^{N\times T_x}$ and the vector $\bar{\mathbf{g}}_k\in\mathbb{C}^{1\times N}$  are both  modeled as  Rayleigh   fading channels, whose each element is an independent and identically distributed (i.i.d.) Gaussian random variable with $\sim\mathcal{CN}(0,1)$. 
	In addition, the RIS architecture that is equipped with additional power circuitry to operate as an active RIS \citep{zhang2021active}, is represented in a diagonal matrix $\mathbf{\Psi}\in\mathbb{C}^{N\times N}=\mathrm{diag}\left\lbrace \alpha_1e^{j\phi_1},  \alpha_2e^{j\phi_2}, \cdots, \alpha_Ne^{j\phi_N} \right\rbrace $, where  $\alpha_n$ and $\phi_n\in[-\pi,\pi]$ being the amplitude and phase of the $n$-th reflecting element for $n\in\left\lbrace 1,2, \cdots, N\right\rbrace $. It is worth noting that since active reflecting elements are capable to amplify the incident signal, the magnitude of each reflecting element is greater than unity, i.e., $\alpha_n>1$.  
	Therefore, for $\bar{T}_x=T_x/K$ being the number of the transmit antennas allocated to each user and $\mathbf{x}_k\in\mathbb{C}^{\bar{T}_x\times1}$ being the  signal vector to be transmitted to the $k$-th user,    the received 	signal at  $\text{U}_k$ is obtained as
	\begin{equation}
		y_k=\sqrt{P_{k}}\mathbf{g}_k\mathbf{\Psi Hx}+\mathbf{g}_k \mathbf{\Psi v}^{\mathrm{T}}+{n}_k
		\label{received1}
	\end{equation}
	where $\mathbf{x}\in\mathbb{C}^{T_x\times1}=[\mathbf{x}_1^{\mathrm{T}}, \mathbf{x}_2^{\mathrm{T}}, \cdots, \mathbf{x}_K^{\mathrm{T}}]^{\mathrm{T}}$ and $\mathbb{E}\left\lbrace \mathbf{x}^{\mathbf{H}} \mathbf{x}\right\rbrace=1 $. Here, $P_k$ is the transmit power dissipated to the $k$-th user, the vector  $\mathbf{v}\in\mathbb{C}^{1\times N}$ represents the  thermal noise  generated from power amplifier circuits of active reflecting elements  \citep{zhang2021active} and ${n}_k$ is the static noise term  at $\text{U}_k$, where $\mathbf{v}\sim\mathcal{CN}(\mathbf{0},\mathbf{I}_N\sigma_v^2)$ and ${n}_k\sim\mathcal{CN}(0,\sigma_s^2)$ for $\sigma_v^2$ and $\sigma_s^2$ being the corresponding noise variances of dynamic and static noise figures, respectively. Moreover, at the user side, since the received superposed signal    at  $\text{U}_k$ (\ref{received1}) includes the targeted and interference signals, it can be rewritten as
	
	\begin{align}
		& y_k=\sqrt{P_{k}}\mathbf{g}_k\mathbf{\Psi H}_k\mathbf{x}_k+\sqrt{P_{k}}\sum_{i\neq k}^{K}\mathbf{g}_k\mathbf{\Psi H}_i\mathbf{x}_i+\mathbf{g}_k \mathbf{\Psi v}^{\mathrm{T}}+{n}_k
		\label{received2}
	\end{align}
	where $\mathbf{H}=[\mathbf{H}_1, \mathbf{H}_2, \cdots, \mathbf{H}_K]$ and $\mathbf{H}_k\in\mathbb{C}^{N\times\bar{T}_x}$ is the channel matrix  between the transmit antenna group dedicated to $k$-th user and the RIS. At this point, the SINR at $\text{U}_k$ can be calculated as:
	\begin{equation}
		\gamma_k=\frac{P_{k}\left\| \mathbf{g}_k\mathbf{\Psi H}_k\right\|^2}{P_{k}\sum_{i\neq k}^{K}\left\| \mathbf{g}_k\mathbf{\Psi H}_i\right\|^2+\left\| \mathbf{g}_k\mathbf{\Psi}\mathbf{v}^{\mathrm{T}}\right\|^2+\sigma_s^2 }.
		\label{sinr}    
	\end{equation}
	Accordingly, the sum-rate of the overall system becomes:
	\begin{equation}
		R=\sum_{k=1}^{K}\log_2(1+\gamma_k).
	\end{equation}
	Then, to maximize this sum-rate, the reflection coefficients of the active RIS elements are optimized. In what follows, the corresponding problem formulation and the proposed solution are presented.
	\subsection{Problem Formulation and Proposed Solution}
	
	In the over-the-air  beamforming-based multi-user transmission scheme,  interference cancellation is performed at the RIS without employing any additional integrated high-cost signal processing circuitry, such as multiple RF chains,  either at T or user sides. {For this purpose,} the reflection coefficients of the RIS are adjusted to maximize the SINR of the intended  $\text{U}_k$. Therefore, to deal with this problem, the following QCQP problem is formulated \vspace{-1cm}
	\begin{equation}
		\nonumber	
	\end{equation}
	\begin{align}
		&	\hspace{-1.1cm}(\text{P1}):\hspace{0.2cm}\max_{\mathbf{\Psi}} \hspace{0.3cm} \gamma_k\label{P1}\\	
		&\hspace{0.3cm} \text{s.t.} \hspace{0.4cm}  {P_{k}\left\| \mathbf{g}_k\mathbf{\Psi H}_k\right\|^2}\geq\Gamma_k\Big( {P_{k}\sum_{i\neq k}^{K}\left\| \mathbf{g}_k\mathbf{\Psi H}_i\right\|^2+\left\| \mathbf{g}_k\mathbf{\Psi}\mathbf{v}^{\mathrm{T}}\right\|^2+\sigma_s^2 }\Big)\label{C1}\\
		&  \hspace{1.20cm} P_{\text{BS}}\left\|\mathbf{\Psi H} \right\|^2 +\left\|\mathbf{\Psi} \right\|^2\sigma_s^2\leq P_A       
		\label{C2}
	\end{align}
	where $\Gamma_k$ is the minimum SINR requirement of $\text{U}_k$,  $P_{\text{BS}}=KP_k$, and $P_A$ is the maximum reflection power introduced by the active reflecting elements.  {Please note that   for the over-the-air beamforming-based multi-user systems, the total power $P_T$ is the  sum of power dissipated   at the transmitter ($P_{\text{BS}}$) and the RIS ($P_A$), that is $P_T=P_A+P_{\text{BS}}$,  while  { for the conventional   transmission without RIS},   $P_T$ denotes to total power consumed at the transmitter.}	
	Then, using the Cauchy-Schwarz inequality, the constraint in (\ref{C2}) can be rewritten as
	\begin{equation}
		\left\|\mathbf{\Psi} \right\|^2\leq\frac{P_A}{P_{\text{BS}}\left\|\mathbf{ H} \right\|^2 +\sigma_s^2}. 
		\label{C3}
	\end{equation}
	Therefore, since   the problem (P1) is  non-convex and it is difficult to obtain an optimal solution,  we resort to  the SDR technique and define new variables ${\mathbf{A}}_k\in\mathbb{C}^{N\times N}=\mathbf{ H}_k\mathbf{H}_k^{\mathrm{H}}$, $\mathbf{B}_k\in\mathbb{C}^{N\times N}=\mathbf{g_k}^{\mathrm{H}}\mathbf{g}_k$ and $\mathbf{V}=\mathbf{vv}^{\mathbf{H}}$. In light of these,  the SINR of the $k$-th user in (\ref{sinr}) can be rewritten as 
	\begin{equation}
		\gamma_k=\frac{P_k\mathrm{Tr}(\mathbf{Q}_k\mathbf{Z})}{ \sum_{i\neq k}P_k\mathrm{Tr}(\mathbf{Q}_i\mathbf{Z})+\mathrm{Tr}(\mathbf{Q}_m\mathbf{Z})+\sigma_s^2}\geq\Gamma_k		
		\label{sinr2}
	\end{equation}
	where  $\mathbf{Q}_k\in\mathbb{C}^{N\times N}=\mathbf{A}_k\circ\mathbf{B}_k$ , $\mathbf{Q}_i\in\mathbb{C}^{N\times N}=\mathbf{A}_i\circ\mathbf{B}_k$ and $\mathbf{Q}_m\in\mathbb{C}^{N\times N}=\mathbf{V}_k\circ\mathbf{B}_k$, while  $\mathbf{Z}\in\mathbb{C}^{N\times N}=\mathbf{z}\mathbf{z}^{\mathrm{H}} $   for   $\mathbf{z}\in\mathbb{C}^{N\times 1}$ being a vector consisting the non-zero diagonal elements of the reflection matrix $\mathbf{\Psi}$, i.e., $\mathbf{z}=[\alpha_1e^{j\phi_1},   \alpha_2e^{j\phi_2}, \cdots, \alpha_Ne^{j\phi_N}]^{\mathrm{H}}$ \citep{zhang2017matrix, ye2020joint}.
	Therefore, the maximization problem (P1) is 	equivalently defined as \vspace{-1cm}
	\begin{equation}
		\nonumber
	\end{equation}
	\begin{align}
		& \hspace{-1.1cm}(\text{P2}):\hspace{0.2cm} \max_{\mathbf{\Psi}} \hspace{0.3cm} \gamma_k\label{P2}\\
		& \hspace{0.25cm} \text{s.t.}\hspace{0.4cm} P_k\mathrm{Tr}(\mathbf{Q}_k\mathbf{Z})-\Gamma_k\big( \sum_{i\neq k}P_k\mathrm{Tr}(\mathbf{Q}_i\mathbf{Z})+\mathrm{Tr}(\mathbf{Q}_m\mathbf{Z})+\sigma_s^2\big)\geq0 \label{C4}\\
		&  \hspace{1.15cm} \mathrm{Tr}(\mathbf{Z}) \leq{\frac{P_A}{P_{\text{BS}}\mathrm{Tr}(\mathbf{ HH}^{\mathrm{H}}
				) +\sigma_s^2}   }.    
		\label{C5}
	\end{align}
	Here, $\mathbf{Z}$ is a positive semidefine matrix and $\mathrm{rank}(\mathbf{Z})=1$. However, since the rank-one constraint is non-convex, we remove this constraint and  reformulate (P2) as a convex feasibility problem as follows
	\vspace{-1cm}
	\begin{equation}
		\nonumber
	\end{equation}
	\begin{align}
		&\hspace{-1.2cm}(\text{P3}):\hspace{0.2cm}\text{Find}\hspace{0.2cm} \mathbf{Z}  \\
		& \hspace{0.2cm}\text{s.t.}  \hspace{0.3cm} P_k\mathrm{Tr}(\mathbf{Q}_k\mathbf{Z})-\Gamma_k\big( \sum_{i\neq k}^K P_k\mathrm{Tr}(\mathbf{Q}_i\mathbf{Z})+\mathrm{Tr}(\mathbf{Q}_m\mathbf{Z})+\sigma_s^2\big)\geq0\label{c1}\\
		&\hspace{1.1cm}\mathrm{Tr}(\mathbf{Z}) \leq{\frac{P_A}{P_{\text{BS}}\mathrm{Tr}(\mathbf{ HH}^{\mathrm{H}}
				) +\sigma_s^2}}\label{c2}.    
	\end{align} 
	Finally,   through  the existing solvers of  CVX toolbox \citep{gb08},  a feasible solution of (P3) satisfying the inequality constraints in  (\ref{c1}) and (\ref{c2}) is obtained.   However, after the relaxation, the optimal solution of (P3) cannot always ensure the rank-one solution. Therefore,  for  $\tilde{\mathbf{Z}}$ being the optimal solution of the problem (P3), {using the eigenvalue decomposition of  $\tilde{\mathbf{Z}}=\mathbf{U\Sigma U}^{\mathrm{H}}$,} the estimated $\mathbf{z}$ is  sub-optimally   obtained  as
	\begin{equation}
		\tilde{\mathbf{z}}=\mathbf{U\Sigma}^{1/2}\mathbf{e}^{\mathrm{H}}
		\label{zhat}
	\end{equation}
 where $\mathbf{e}\in\mathbb{C}^{1\times N}$ is a Gaussian random vector with  $\sim\mathcal{CN}(\mathbf{0}, {\mathbf{I}})$, where $\mathbf{U}\in\mathbb{C}^{N\times N}$ is a unitary matrix of eigenvectors and $\mathbf{\Sigma}\in\mathbb{C}^{N\times N}$ is a diagonal matrix of eigenvalues. {Then, after determining optimized reflection matrix, the RIS performs over-the-air beamforming in order to alleviate the user interference. }
	

	\section{ Over-the-Air  Receive Index Modulation} 
	In this section, the proposed over-the-air beamforming concept is adopted to a novel receive IM transmission scheme. Considering the over-the-air beamforming approach given in Section II, a single-user uplink transmission of an active RIS-aided IM transmission system is developed.
	
	\subsection{System Model of Over-the-Air Receive IM}
	\begin{figure}[h!]
		\begin{center}
			\includegraphics[width=0.75\linewidth]{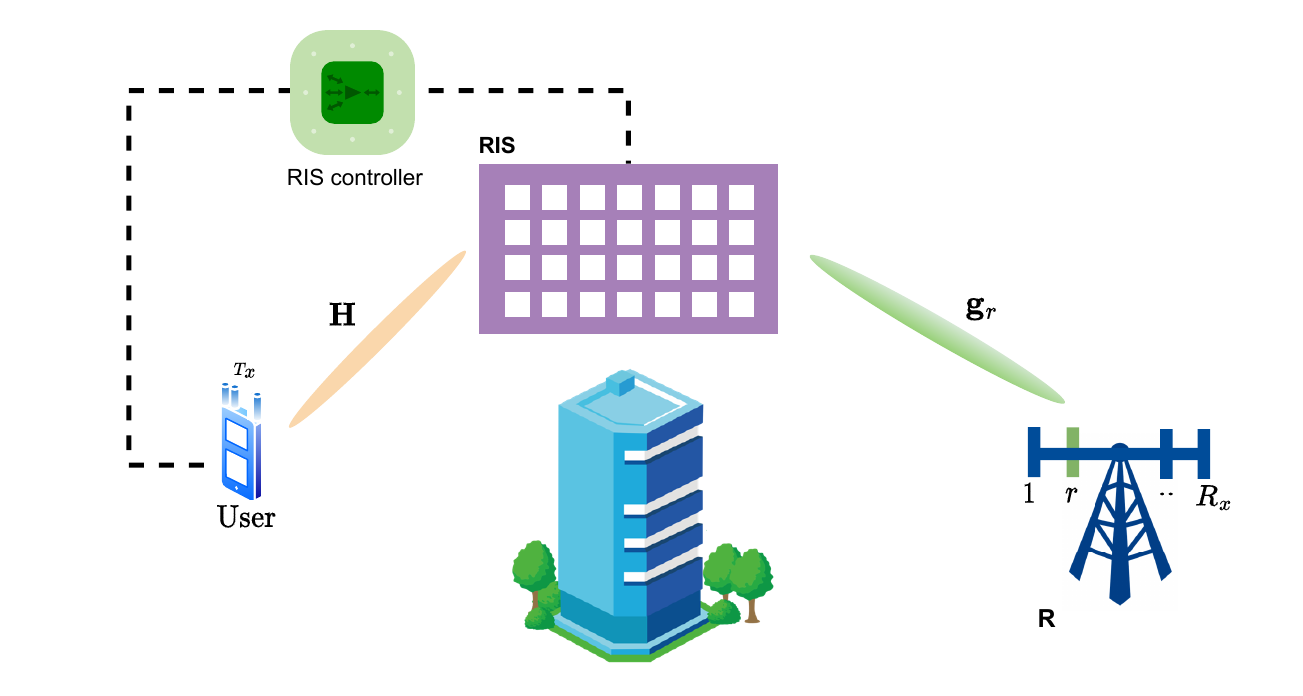}
		\end{center}
		\caption{ Over-the-air  receive IM scheme. }\label{sys2}
	\end{figure}

	As given in  Figure  \ref{sys2}, in the proposed IM system, due to presence of the obstacles over the direct links, a multi-antenna user communicates with an $R_x$-antenna receiver (R) through an RIS with $N$ reflecting elements.  Besides, an RIS controller is attached to the RIS that exchanges the information through a feedback control link.  In the proposed system, considering  the IM transmission principle \citep{basar2020reconfigurable}, an over-the-air receive IM scheme is developed. Unlike traditional  receive IM schemes \citep{stavridis2012transmit,zhang2013generalised,luo2021spatial} that deploy transmit precoding techniques   via  high-cost hardware devices for preprocessing the transmit signal before its transmission, { the proposed receive  IM scheme employs the RIS as a signal processing  unit and apply an over-the-air beamforming at the RIS}.   In the over-the-air receive IM scheme,  at the user side, the  conventional multi-antenna transmission  is considered. Moreover, in order to attain higher data rates, extra information bits are conveyed via  indicating the active receive antenna index.  Therefore,  the incoming  information bits are used to determine the modulated $M$-PSK symbols for  each of the available $T_x$   transmit antennas, as well as to specify the active receive antenna index, one   out of $R_x$ receive antennas. Therefore, the spectral efficiency achieved by this  novel receive IM scheme is calculated as
	\begin{equation}
		\eta_{\text{IM}}=T_x\log_2(M)+\log_2(R_x) \hspace{1cm} \text{[bits/s/Hz]}.
		\label{eta_IM}
	\end{equation} In this system, the information of the active receive antenna index and  perfect channel knowledge of  {user}-RIS and RIS-R links is shared by the user to the  RIS through the smart  controller. Then, the  reflection coefficient of the RIS elements are adjusted to ensure that the target receive antenna has the strongest received signal power. In other words, by the means of active reflecting elements, the RIS acts  as a kind of digital beamformer  and steers the overall signal along the  desired receive antenna direction. 
	
	Let the multi-path fading channels between user-RIS and RIS-R links are modeled as the independent Rayleigh fading channels, which are denoted by the channel matrices  of  {$\mathbf{H}\in\mathbb{C}^{N\times T_x}$}  and  {$\mathbf{G}\in\mathbb{C}^{R_x\times N}=[\mathbf{g}_1^{\mathrm{T}},\mathbf{g}_2^{\mathrm{T}}, \cdots, \mathbf{g}_{R_x}^{\mathrm{T}}]^{\mathrm{T}}$}, respectively, where  $\mathbf{g}_{r}\in\mathbb{C}^{1\times N}$  is the  $r$-th  row of the the channel matrix $\mathbf{G}$  corresponding to the  channel vector between the RIS and the $r$-th receive antenna for $r\in\left\lbrace 1,2,\cdots, R_x\right\rbrace $. 
	Therefore, for  $x_t$  being the  $M$-PSK modulated signal transmitted from the $t$-th transmit antenna, the overall transmit signal becomes $\mathbf{x}\in\mathbb{C}^{T_x\times1 }=[x_1,\cdots, x_{T_x}]^{\mathrm{T}}$,  where $\mathbb{E}\left\lbrace \mathbf{x}^{\mathrm{H}}\mathbf{x} \right\rbrace=1 $ and $t\in\left\lbrace1,2,\cdots, T_x \right\rbrace $. Then, the received signal at the target receive antenna $r$  is obtained as
	\begin{equation}
		y_r=\sqrt{P_{\text{BS}}}\mathbf{g}_r\mathbf{\Psi}_r\mathbf{ Hx}+\mathbf{g}_r\mathbf{\Psi}_r\mathbf{v}+{{n}_r}
		\label{receive_im}
	\end{equation}
	where $\mathbf{\Psi}_r\in\mathbb{C}^{N\times N}$ is the optimized diagonal reflection matrix for the corresponding $r$-th receive antenna. 
	It is worth noting that   according to incoming spatial bits, if the $r$-th receive antenna is    activated,  it is ensured that the signal power  of the  $r$-th   received  antenna is much stronger than the others:  
	\begin{equation}
		\left\| \mathbf{g}_r\mathbf{\Psi}_r \mathbf{H}\right\|^2\gg \sum_{i\neq r}^{R_x}\left\| \mathbf{g}_i\mathbf{\Psi}_r \mathbf{H}\right\|^2.
		\label{karsi}
	\end{equation}
	Therefore,  to address this problem, for $\mathbf{\Theta}_r=\mathrm{diag}(\mathbf{g}_r)\mathbf{H}$ and $\mathbf{z}\in\mathbb{C}^{1\times T_x}=\mathrm{diag}(\mathbf{\Psi}_r)$, a QCQP  optimization problem is formulated as\vspace{-1cm}
	\begin{equation}
		\nonumber
	\end{equation}
	\begin{align}
		& \hspace{-1.1cm}(\text{P4}):\hspace{0.1cm}\max_{\mathbf{\Psi}}\hspace{0.1cm}\mathbf{z}^{\mathrm{H}}\mathbf{\Theta}_r\mathbf{\Theta}_r^{\mathrm{H}}\mathbf{z}\label{P4}\\
		& \hspace{0.25cm} \text{s.t.}\hspace{0.4cm}  P_{\text{BS}}\left\|\mathbf{\Psi}_r\mathbf{ H} \right\|^2 +\left\|\mathbf{\Psi}_r \right\|^2\sigma_s^2\leq P_A.        
		\label{CIM}
	\end{align}
	Then, resorting to   SDR,    the problem (P4) is expressed as\vspace{-1cm}
	\begin{equation}
		\nonumber
	\end{equation}
	\begin{align}
		& \hspace{-1.1cm}(\text{P5}):\hspace{0.1cm}\max_{\mathbf{\Psi}}\hspace{0.1cm}\mathrm{Tr}(\mathbf{\Delta}_r\mathbf{Z})\label{P4}\\
		& \hspace{0.25cm} \text{s.t.}\hspace{0.4cm} \mathrm{Tr}(\mathbf{\Delta}_r\mathbf{Z})- \delta_r\sum_{i\neq r}^{R_x}\mathrm{Tr}(\mathbf{\Delta}_i\mathbf{Z})\geq0\\
		& \hspace{1.15cm}     \mathrm{Tr}(\mathbf{Z}) \leq{\frac{P_A}{P_{\text{BS}}\mathrm{Tr}(\mathbf{ HH}^{\mathrm{H}}
				) +\sigma_s^2}}.
		\label{CIM}
	\end{align} 
	Here, for  $\delta_r\gg1$, $\mathbf{\Delta}_r\in\mathbb{C}^{N\times N}=\mathbf{\Theta}_r\mathbf{\Theta}_r^{\mathrm{H}}$ and $\mathbf{Z}=\mathbf{z}\mathbf{z}^{\mathrm{H}}$, the problem (P5) is solved using   CVX  solvers \citep{gb08}. Then, following the same processes as in the  multi-user downlink transmission in Section II, the sub-optimal  estimate of  $\mathbf{z}$, is  obtained as given in   (\ref{zhat}). {Then, the resulting RIS reflection matrix enables that  the overall signal is oriented in the direction of the target receive antenna.}
	\subsection{Low-Complexity Successive Greedy Detector }
	In the subsection that follows,    a sub-optimal successive detection algorithm for the proposed receive IM scheme is proposed. 
		In the proposed system,   after the optimization of the reflection matrix $\mathbf{\Psi}_r$ for the specified $r$-th receive antenna,  it is straightforward to exploit a  maximum likelihood (ML) detector that jointly estimates the "spatial symbol"  $r$ and the overall transmit signal vector $\mathbf{x}$ as follows
	\begin{equation}
		[\hat{r},\hat{\mathbf{x}}]=\arg\min_{\substack{r, \mathbf{x}}}\sum_{j=1}^{R_x}\Big| {y}_j-\sqrt{P_{\text{BS}}}\mathbf{g}_j\mathbf{\Psi}_r\mathbf{ Hx}\Big|^2.
	\end{equation}
	
	However, in the proposed receive IM scheme, in order to save the computational complexity,  instead of considering joint detection, the  receiver reconstructs the transmit information via a low-complexity greedy detector that perform the successive detection  in the following way.  First,   using  amplitude detectors, the index of the active receive antenna is detected as
	\begin{equation}
		\hat{r}=\arg\max_{j\in\left\lbrace1, \cdot\cdot,R_x \right\rbrace }\left|y_j \right|. 
	\end{equation}
	Then,  exploiting the  maximum likelihood (ML) detector, the transmit signal vector $\mathbf{x}$ is estimated, by considering all possible $\mathbf{x}$ realizations, as follows
	\begin{equation}	
		\hat{\mathbf{x}}=\arg\min_{\mathbf{x}}\Big| y_{\hat{r}}-\sqrt{P_{\text{BS}}}\mathbf{g}_r\mathbf{\Psi}_r \mathbf{Hx}\Big|^2. 
	\end{equation}
	
	Moreover, from the computational complexity standpoint,  we note that since the complexity of SDR problem (P5) is $\mathcal{O}(N^{4.5})$ \citep{luo2010semidefinite}, the overall complexity of the greedy detector approximates to  $\sim\mathcal{O}(M^{T_x}+T_x)$, while the complexity  for the {joint} ML detector  is  $\sim\mathcal{O}((M^{T_x}+T_x+N^{4.5})R_x^2)$, which grows exponentially with increasing $N$ and $R_x$. {Therefore,  comparing to the joint ML detection, the proposed greedy detector offers a significant  reduction in computational burden.} 
	
	
	\section{Numerical Results}
	In this section, the sum-rate and BER performance of the proposed over-the-air beamforming-based single-user and multi-user   downlink transmission, and uplink receive IM schemes are presented through the Monte Carlo simulations. Moreover,  comparing to the  ZF-based conventional transmission  \citep{spencer2004zero, zhang2013generalised} and the state-of-the-art RIS-aided joint beamforming schemes \citep{wu2019intelligent}, the improved performance of the over-the-air beamforming-based systems are illustrated.

	In all computer simulations, the following system setups are considered: the  reference path loss value is $C_0=-30$ dBm, the noise variances are $\sigma_v^2=\sigma_s^2=-90$ dBm , the path loss exponents for the RIS-aided systems are $\beta_T=2.2$ and  $\beta_k=2.8$ and for the conventional direct transmission, it is   $\beta_D=3.5$ \citep{nguyen2022hybrid}, the distances are $d_T=20$,  $d_R=30$ m and $d_{D}=50$ m. 
		\begin{figure}[htp!]
		\begin{center}
			\includegraphics[width=0.67\linewidth]{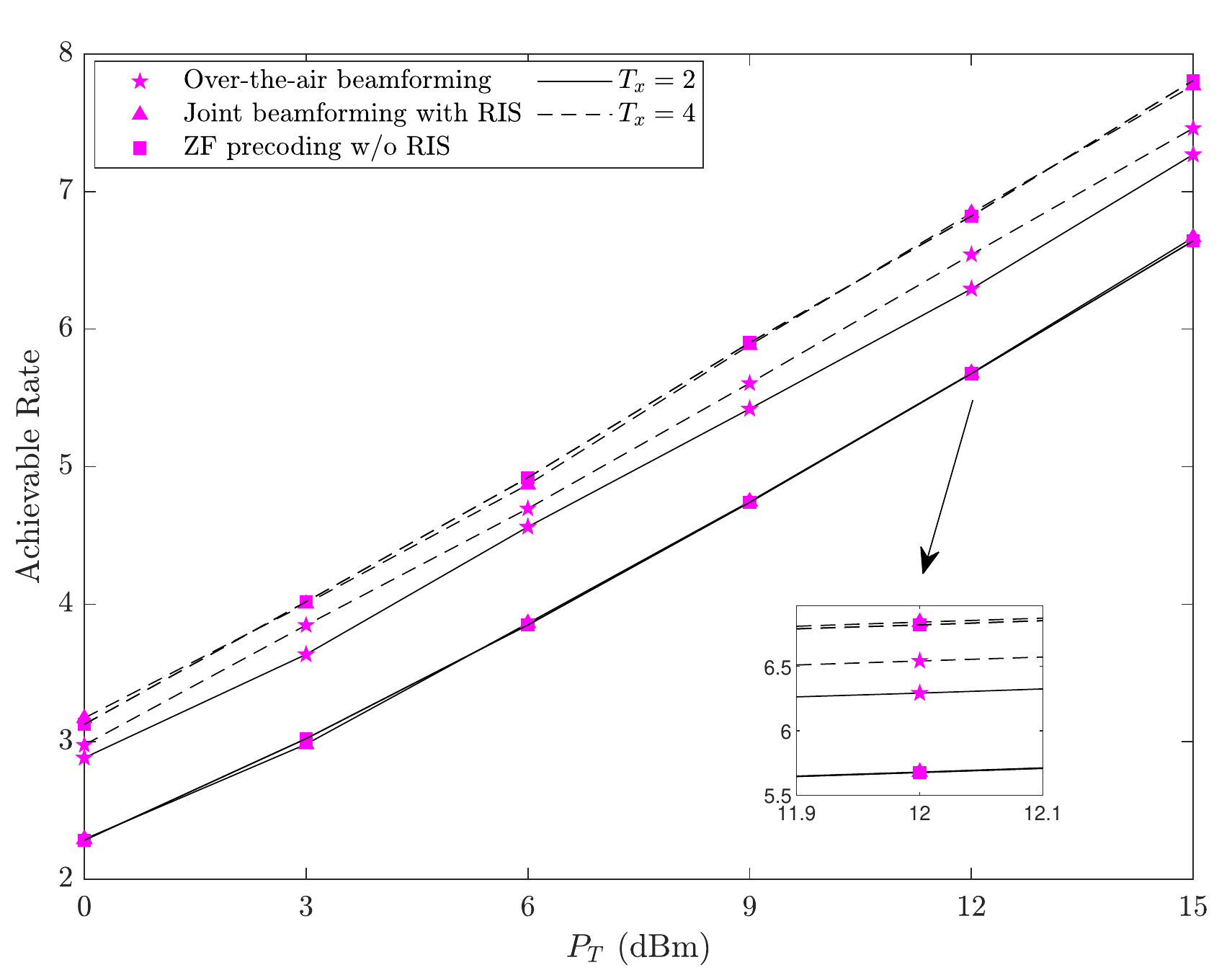}
		\end{center}
		\caption{ Comparison of the achievable rate performance  of the proposed over-the-air beamforming  with traditional ZF precoding \citep{spencer2004zero} and joint beamforming with RIS \citep{wu2019intelligent} for  single-user system configurations.}\label{SU}
	\end{figure}

	\subsection{Downlink transmission}
	In this subsection, the {numerical  } results of the proposed over-the-air beamforming and the  benckmark schemes for single-user and multi-user   downlink systems  are  demonstrated. 
	
	\subsubsection{Single-user}
	{The following  computer simulation results are performed for single-user MISO transmission schemes.

 In Figure \ref{SU}, for a single-user  downlink transmission ($K=1$) with $T_x\in\left\lbrace2,4 \right\rbrace $ and $N=16$, the achievable rate performance of the proposed over-the-air beamforming scheme   as a function of total transmit power $P_T$ is compared to the traditional ZF precoding \citep{spencer2004zero} and the passive RIS-aided joint active and passive beamforming techniques \citep{wu2019intelligent}. Here,   while $P_T$ is the  overall power consumed at the transmitters of the traditional ZF precoding and joint beamforming transmission schemes,  it corresponds to the total power dissipated between the transmitter ($P_{\text{BS}}$) and the RIS ($P_A$) for the proposed over-the-air transmission, where $P_T=P_{\text{BS}}+P_A$ and $P_{\text{BS}}=0$ dBm. Moreover, as discussed in Section 2.1,  the reference ZF precoding considers a traditional single-hop transmission without RIS that performs transmit precoding before the signal transmission \citep{spencer2004zero}. On the other hand, in  the joint active and passive beamforming scheme, a passive RIS-aided   single-user transmission with the existence of  direct links between the transmitter and the user, is considered, where  the digital beamforming at the transmitter (active) and analog beamforming at a passive RIS via phase shifters are jointly optimized to enhance the received SNR of the user \citep{wu2019intelligent}. For this purpose, similar to our proposed beamforming technique, a QCQP-based non-convex optimization problem is formulated and an SDR-based solution is performed via CVX solvers \citep{wu2019intelligent}.    The results show that  although a direct link between the transmitter and the user does not exist in the proposed active RIS-aided over-the-air beamforming scheme, a considerably better performance achievement is observed  for $T_x=2$ compared to   the traditional ZF and joint beamforming with passive RIS-aided transmission  schemes. Moreover, it is shown that  increasing $T_x$ results in enhancement of the achievable rate  of all systems. However, in the proposed active RIS-aided over-the-air beamforming scheme,  as given in (\ref{C3}), since the magnitude of reflection matrix $\mathbf{\Psi}$  is restricted with  the magnitude of transmission matrix $\mathbf{H}$, i.e., increasing $T_x$,    a slighter performance improvement is achieved compared to the benchmark schemes. Furthermore,  since a small-scale passive RIS is considered, i.e. $N=16$, an additional performance improvement   due to the  indirect RIS-aided link is hardly observed in the passive RIS-aided joint beamforming scheme compared to the conventional ZF precoding scheme. }   
		\begin{figure}[t]
		\begin{center}
			\includegraphics[width=0.69\linewidth]{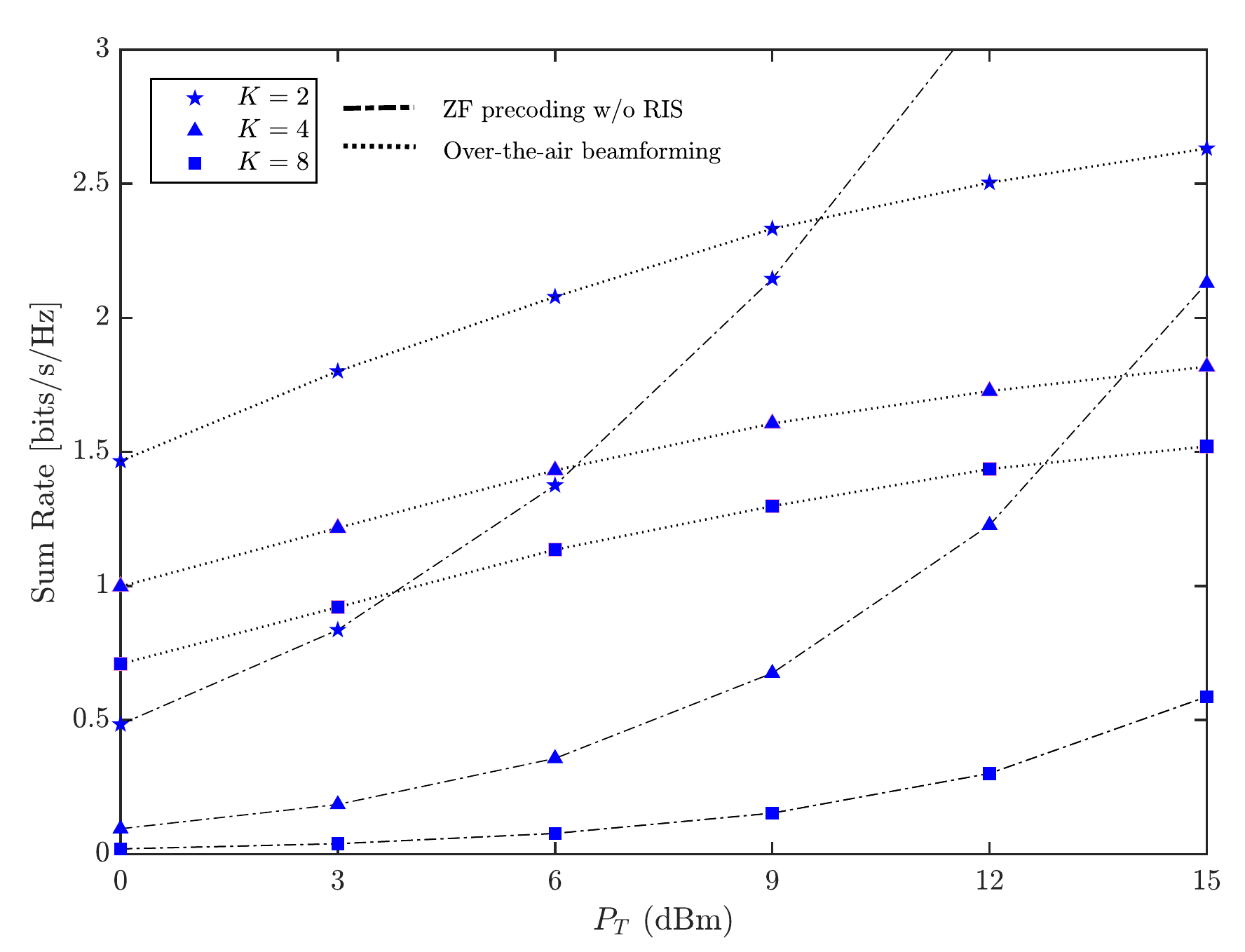}
		\end{center}
		\caption{ Sum-rate  comparison of the proposed over-the-air beamforming and the classical transmit ZF  precoding for   $K\in\left\lbrace2,4,8 \right\rbrace $.}\label{ZF_kars}
	\end{figure}
	\subsubsection{Multi-user}
	The following results are carried out for downlink multi-antenna transmission schemes, where a single transmit antenna is allocated to each user, i.e., $\bar{T}_x=1$ and $T_x=K$.
	
	In Figure \ref{ZF_kars}, the sum rate of the downlink multi-antenna  transmission scheme based on the conventional transmit ZF  precoding {\citep{spencer2004zero}} and   the novel over-the-air beamforming  has been carried out for $K\in\left\lbrace 2,4,8\right\rbrace $,  and quadrature PSK (QPSK), i.e., $M=4$.  Comparing these two schemes,  it is obvious  that at  lower $P_T$ values, the  over-the-air beamforming based multi-user transmission scheme attains higher sum-rate  than the classical transmit ZF  precoding technique \citep{spencer2004zero}. However, for $K=2$ and $K=4$, as  $P_T$ increases, the performance of ZF gradually begins to exceed the performance of the proposed beamforming scheme.  Nevertheless, for $K=4$, the ZF precoder achieves only a slight gain over the proposed beamforming concept at $P_T=15$ dBm.  It can be also deduced from Figure \ref{ZF_kars} that an increase in  the total number of users rapidly decreases ZF sum-rate, however, such a {severe} performance loss is not observed in the proposed over-the-air beamforming-based system. Moreover, when the number of users  further increases to $K=8$, it is observed that the system with  the proposed over-the-air  beamforming-based scheme outperforms   the system with the traditional ZF technique with a significant performance gain.

\begin{figure}[t]
		\begin{center}
			\includegraphics[width=0.67\linewidth]{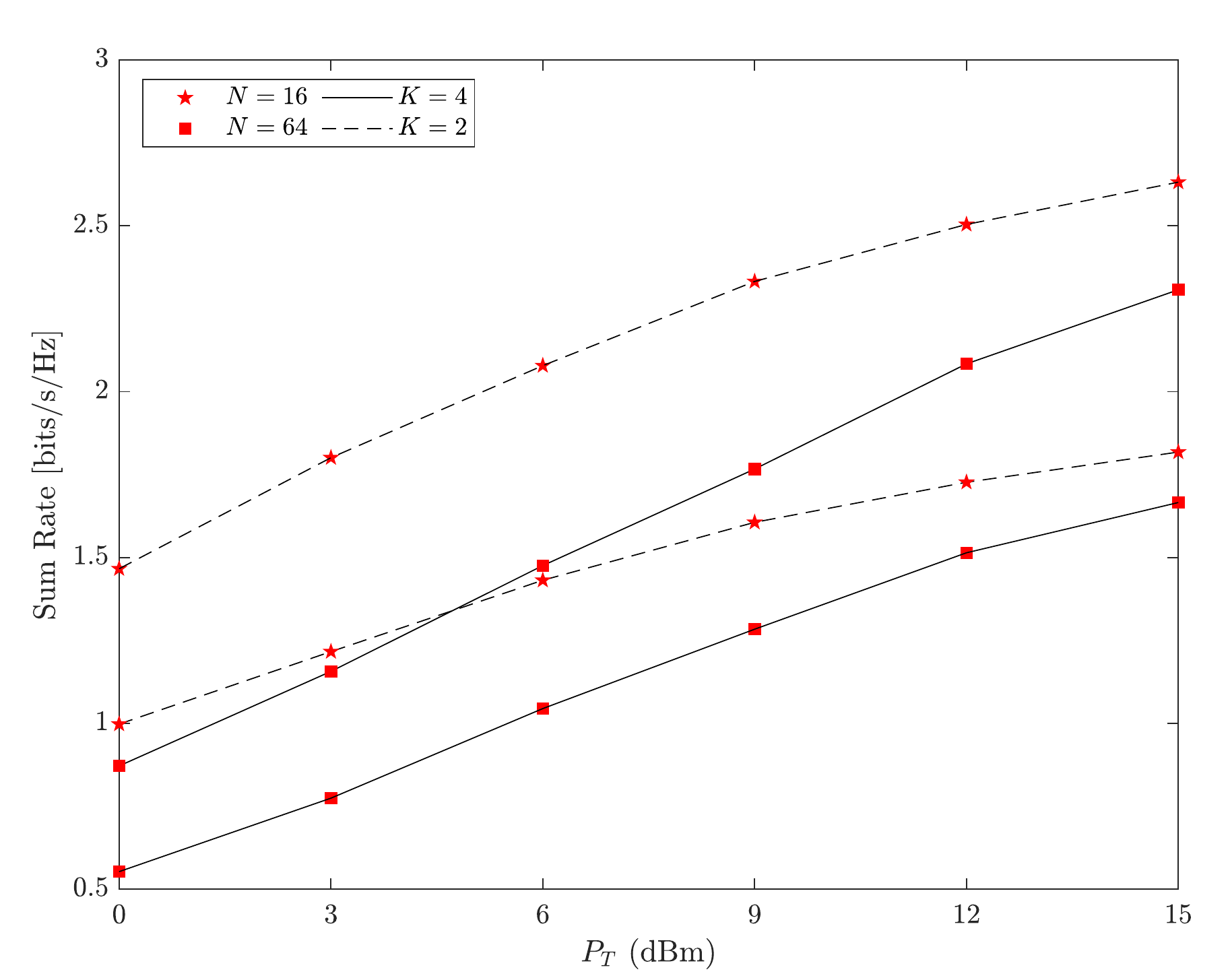}
		\end{center}
		\caption{ Sum-rate performance  of the proposed over-the-air beamforming-based multi-user systems   for   different system configurations}\label{Ptot}
	\end{figure}

	In Figure \ref{Ptot}, the sum-rate of the proposed over-the-air beamforming-based downlink multi-user system is evaluated for different system configurations. In this case, for a constant $P_T$, the performance of the over-the-air beamforming based systems are investigated for  different number of the reflecting  elements $N$,  $P_{\text{BS}}=0$ dBm and QPSK signaling. It is observed that increasing  RIS size has an adverse affect on the system performance. This results may be explained by the fact that in the over-the-air beamforming design, as given in (\ref{C3}), the power consumed by the reflecting elements is inversely proportional with the magnitude of the channel matrix  $\mathbf{H}$. Therefore, when a constant $P_A$ is considered for $N=16$ and $N=64$, it reveals  that the proposed beamforming-based systems with the lower $N$ values show considerably better performance than the ones with  the higher $N$ values.  
	
	\begin{figure}[t]
		\centering
		\includegraphics[width=0.67\linewidth]{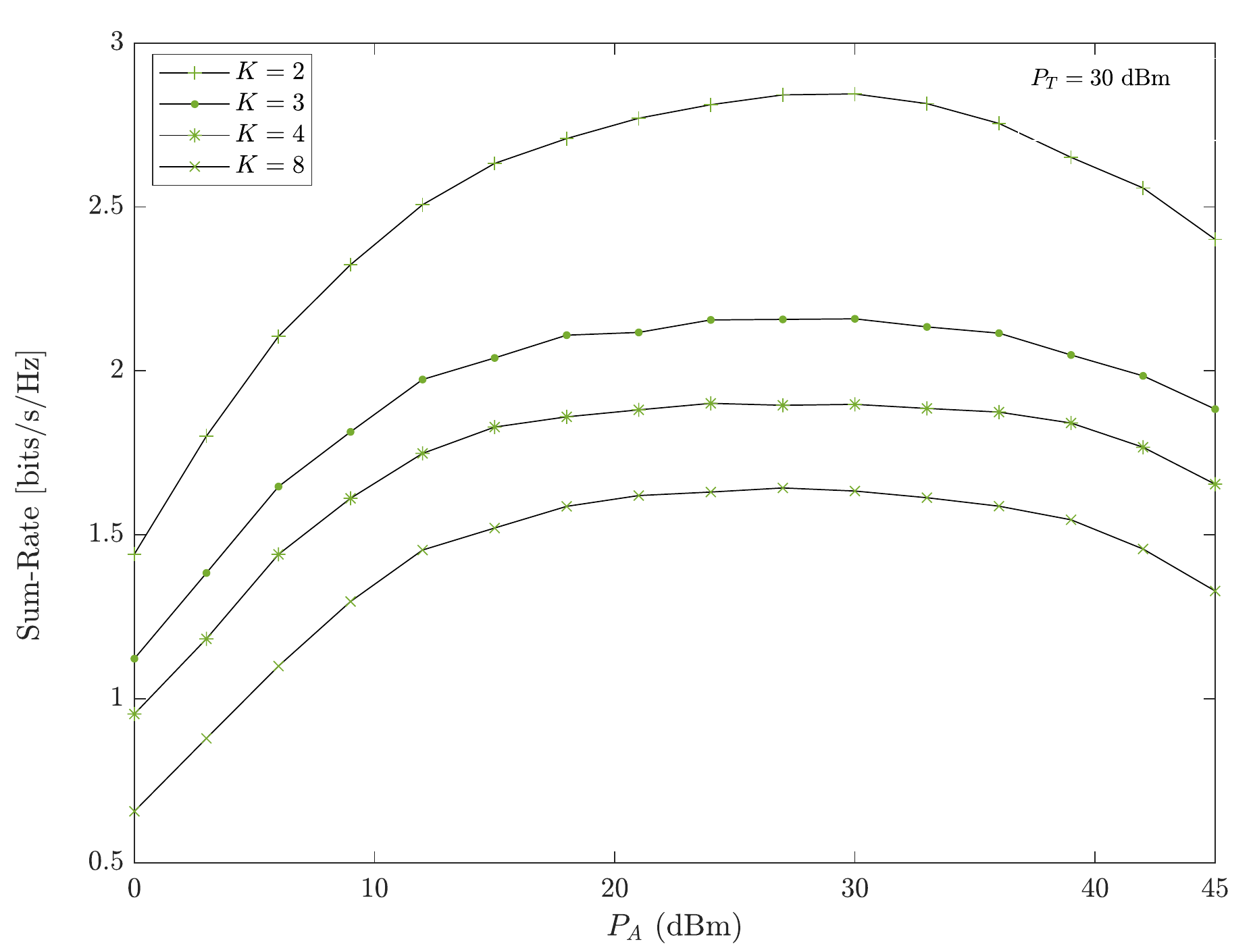}
		\caption{Sum-rate  of the proposed over-the-air beamforming-based systems for  $N=16$ and $K\in\left\lbrace 2,3,4,8\right\rbrace $. }\label{2li}
	\end{figure}

	In Figure \ref{2li}, the effect of increasing reflection power $P_A$ on the sum-rate of the proposed beamforming based systems with QPSK and $P_T=30$ dBm
	is investigated for  $N=16$. The results show that in all cases, the increasing $P_A$ improves the system performance up to a certain $P_A$ value, after which the performance begins to degrade. These results indicates the relation between  the  reflection power constraint $P_A$ and transmitter power $P_{\text{BS}}$ in  (\ref{C2}). Indeed, in our system design,  the overall consumed power $P_T$ is dissipated to the transmitter ($P_{\text{BS}}$) and the RIS   ($P_A$), where $P_T=P_{\text{BS}}+P_A$, and for a constant $P_T=30$ dBm, $P_{\text{BS}}$ decreases with increasing $P_A$. However, it is clear from  (\ref{sinr}) that the {minimizing}  $P_{\text{BS}}$  directly affects the SINR value. Surely,   the investigation of this interesting trade-off points out the importance of the    power allocation between the transmitter and the RIS, which is an  open problem to be addressed in  future studies.
		\begin{figure}[htp!]
		\centering
		\begin{minipage}{.5\textwidth}
			\centering
			\includegraphics[width=.78\linewidth]{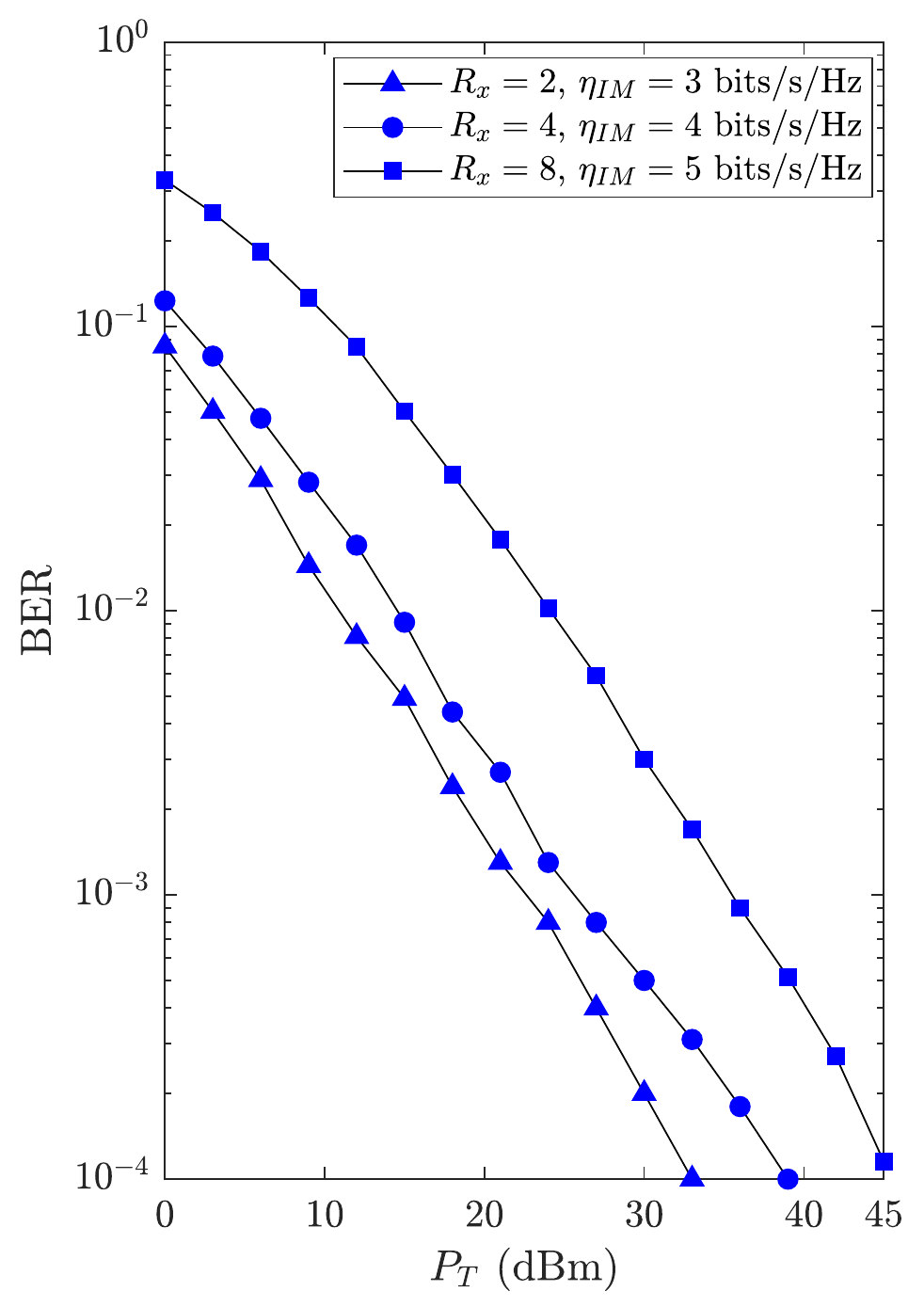}\\
			\hspace{0.2cm}(a)
		\end{minipage}%
		\begin{minipage}{0.5\textwidth}
			\centering
			\vspace*{0.2cm}	\includegraphics[width=.78\linewidth]{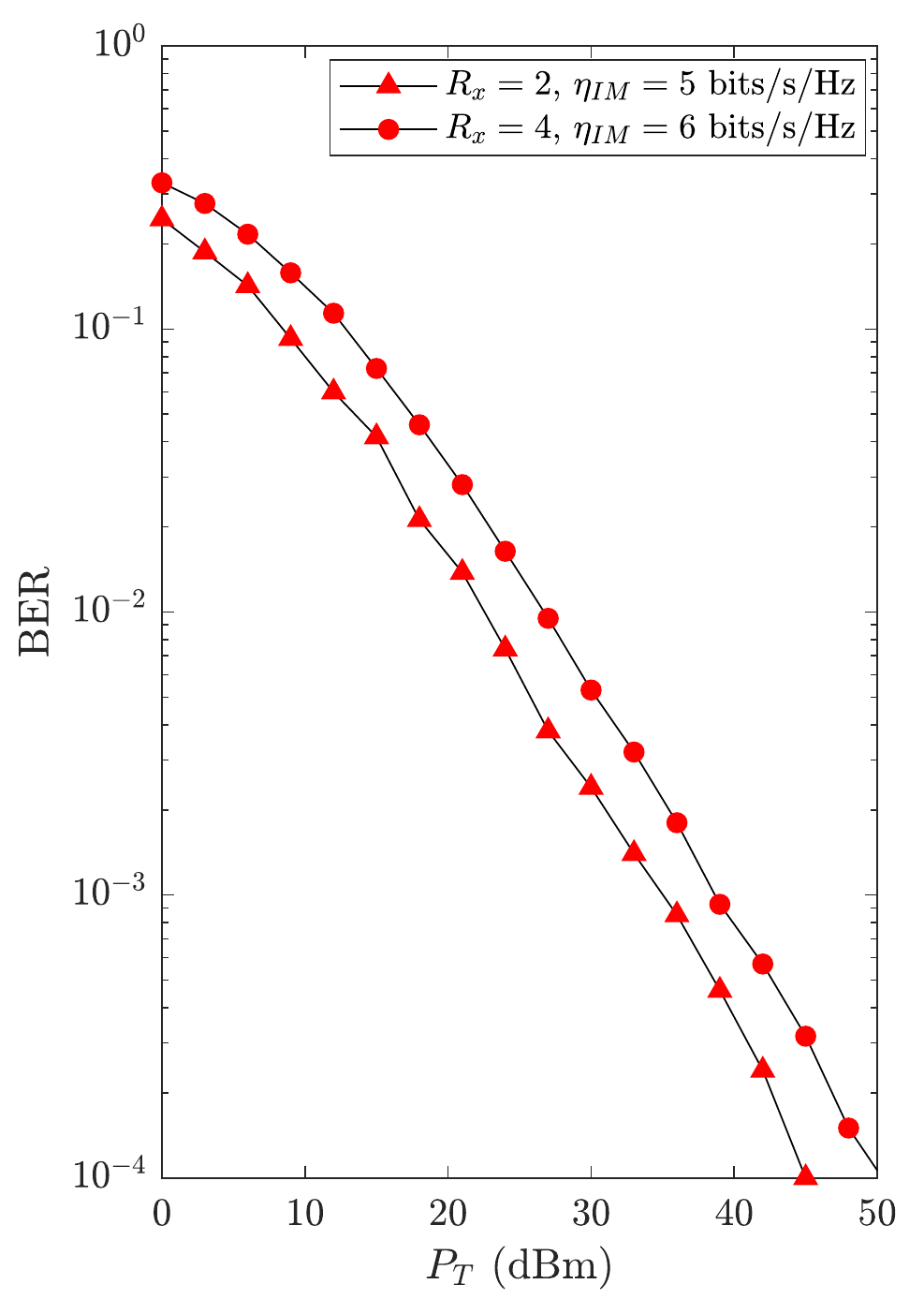}\\
			\hspace{0.2cm}(b)
		\end{minipage}
		\caption{BER performance of the novel receive IM scheme with greedy detector for $N=16$ and  (a) $T_x=2$, (b) $T_x=4$}
		\label{IM}
	\end{figure}
	
	\subsection{Single-user uplink transmission }
	In this subsection, the BER performance of the proposed receive IM scheme is evaluated.

	In Figure \ref{IM}, the BER performance of the proposed receive IM scheme {with sub-optimal greedy detector} is investigated for different  RIS-aided MIMO configurations with $N=16$ and binary PSK (BPSK). 
	Similar to the conventional receive IM schemes \citep{wu2021reconfigurable,zhang2013generalised}, the performance results of  the corresponding  high-rate systems  that employ (a) $T_x=2$ and (b) $T_x=4$ transmit antennas reveal a certain trade-off between system performance and data-rate.

\begin{figure}[!t]
		\centering
		\includegraphics[width=0.73\linewidth]{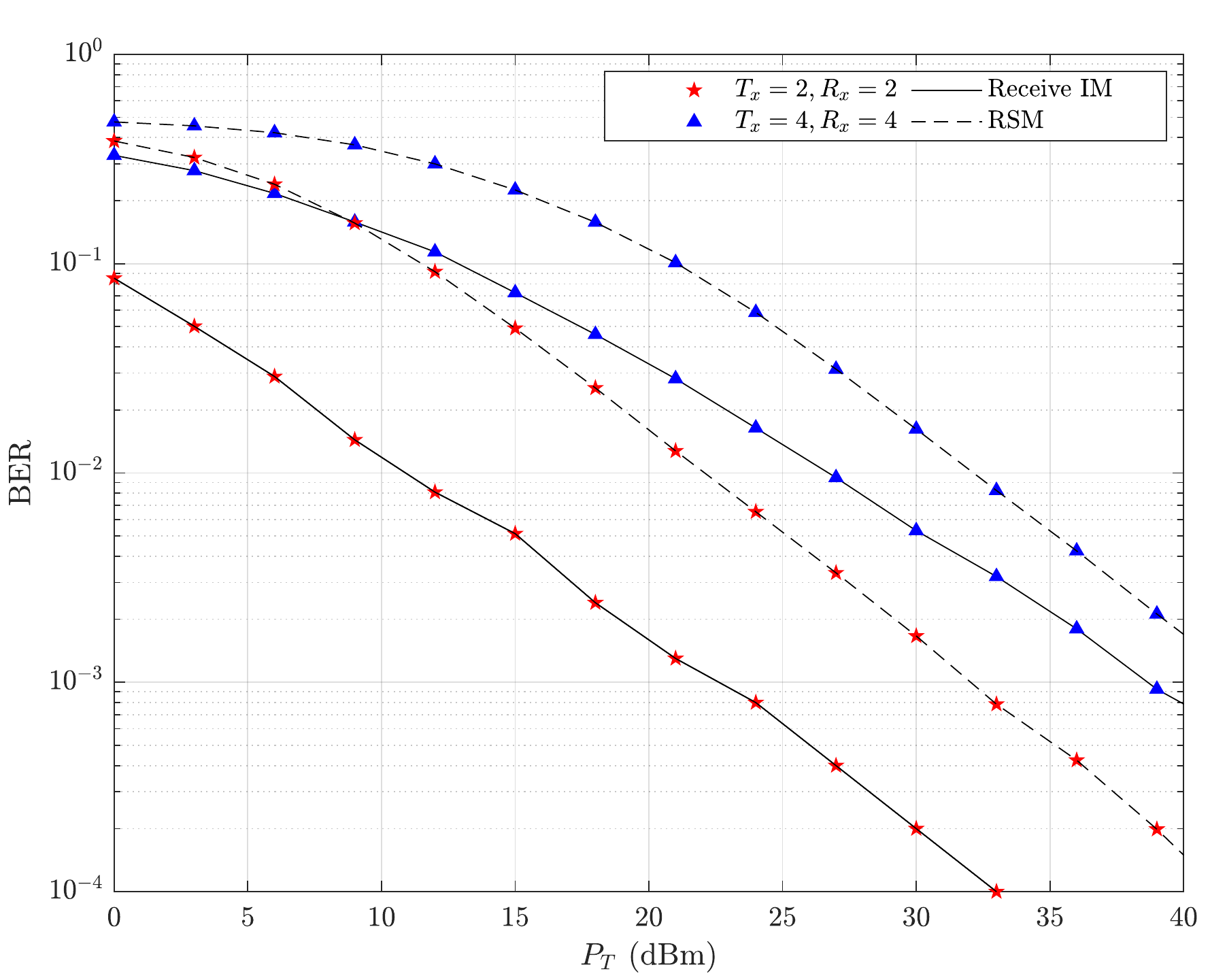}\\
		
		\caption{BER performance comparison of the proposed receive IM and RSM  schemes.}
		\label{IMyeni}
	\end{figure}

	In Figure \ref{IMyeni}, the BER performance of the transmit ZF precoded receive spatial modulation (RSM) \citep{zhang2013generalised} and the proposed over-the-air receive IM schemes are compared. For $R_x=T_x=2$, the receive IM  and the RSM schemes respectively exploit BPSK and  QPSK modulations to achieve $\eta_{\text{IM}}=3$ bits/s/Hz. On the other hand, for $R_x=T_x=4$ configuration, the receive IM with BPSK and the RSM with 16-PSK assess $\eta_{\text{IM}}=6$ bits/s/Hz.   The results demonstrate the significant performance improvement of the proposed receive IM scheme over the traditional ZF precoded RSM \citep{zhang2013generalised}.
	
	\section{Conclusion}
	In this paper, first, deploying an active RIS, a novel  beamforming approach has been proposed for  RIS-aided multi-user systems. In the proposed  concept,   without employing any other signal processing units at the transmitter and/or  receiver sides, the reflection coefficients of the active RIS have been customized to mitigate the user interference. To meet this challenge, we have obtained SDR-based solutions via CVX software toolbox.   Moreover,   taking the  proposed over-the-air beamforming concept one step further, a low-complexity  receive IM  scheme has been developed for  single-user uplink transmission.  Through computer simulations, the enhanced  performance of the over-the-air beamforming-based  systems over the traditional precoding-based systems have been  indicated.

	\bibliographystyle{Frontiers-Harvard} 
	\bibliography{references}
	

	
\end{document}